\documentclass[referee]{aa}

\usepackage{graphicx}
\usepackage{natbib}
\usepackage{latexsym}
\usepackage{amsmath}
\usepackage{amssymb}
\usepackage{amstext}
\usepackage{color}
\usepackage{psfrag}

\def\k{km s$^{-1}$}
\def\ks{km s$^{-1}$~}
\def\d{$^\circ$}
\def\m{$^\prime$}
\def\s{$^{\prime\prime}$}
\def\hh{$^{\mathrm h}$}
\def\mm{$^{\mathrm m}$}
\def\ss{$^{\mathrm s}$}
\def\cm3{cm$^{-3}$}

\def\2{$^{12}$CO}
\def\3{$^{13}$CO}
\def\msol{M$_\odot$}

\begin{document}

\title{Study of the molecular clump associated with the high-energy source HESS J1858+020}

\author {S. Paron \inst{1,2}
 \and E. Giacani \inst{1,2}
 \and M. Rubio \inst{3}
 \and G. Dubner \inst{1}
}
                                                                                                               
\institute{Instituto de Astronom\'{\i}a y F\'{\i}sica del Espacio (CONICET-UBA),
             CC 67, Suc. 28, 1428 Buenos Aires, Argentina\\
             \email{sparon@iafe.uba.ar}
\and FADU - Universidad de Buenos Aires 
\and Departamento de Astronom\'{\i}a, Universidad de Chile, Casilla 36-D,
        Santiago, Chile
}

\offprints{S. Paron}

   \date{Received <date>; Accepted <date>}

\abstract{}{HESS J1858+020 is a weak $\gamma-$ray source lying near the southern border of the SNR G35.6-0.4. 
A molecular cloud, composed by two clumps, shows signs of interaction with the SNR and with a nearby 
extended HII region. In particular, the southernmost clump coincides with the center of the HESS source.
In this work we study this clump in detail with the aim of adding information that helps in the identification
of the nature of the very-high energy emission.} {We observed the mentioned 
molecular clump using the Atacama Submillimeter Telescope Experiment (ASTE) in the \2 J=3--2, \3 J=3--2, 
HCO$^{+}$ J=4--3 and 
CS J=7--6 lines with an angular resolution of 22\s. To complement this observations we analyzed IR and submillimeter
continuum archival data.}{From the \2 and \3 J=3--2 lines and the 1.1 mm continuum emission we derived a density of
between $10^{3}$ and $10{^4}$ cm$^{-3}$ for the clump. We discovered a young stellar object (YSO), probably a high mass protostar, 
embedded in the molecular clump. However, we did not observe any evidence 
of molecular outflows from this YSO which would reveal the presence of a thermal jet capable of generating the 
observed $\gamma$-rays. We conclude that the most probable origin for the TeV $\gamma$-ray emission are 
the hadronic interactions between the molecular gas and the cosmic rays accelerated by the shock front of the SNR G35.6-0.4.}{}

\titlerunning{A molecular clump associated with  HESS J1858+020}
\authorrunning{S. Paron et al.}


\maketitle

\section{Introduction}

HESS J1858+020 is a weak $\gamma-$ray source detected with the Cherenkov telescope
{\it High Energy Stereoscopic System} (HESS). Though nearly point-like
source, its morphology shows a slight extension of $\sim 5^{\prime}$
along its major axis. The source has been detected at a significance level
of 7$\sigma$ with a differential spectral index of 2.2 $\pm$0.1 \citep{aha08}.
The radio source G35.6-0.4, recently identified as a supernova remnant (SNR) by 
\citet{green09}, is seen in projection over the northern border of HESS J1858+020.
The author estimated an age of 30000 years for the SNR and a distance 
of $\sim$ 10.5 kpc. 
Very recently, \citet{yo10} studied the interstellar medium (ISM) around the
very-high energy source and, on the basis of \3 J=1--0 data, they identified a molecular cloud composed by 
two clumps. One of these clumps is seen in projection over the southern border of SNR G35.6-0.4 
and presents some kinematical signatures of disturbed gas, while the other clump 
coincides with the center of HESS J1858+020. Based on an IR study, \citet{yo10} found 
evidence of star formation activity in the second mentioned clump. They suggested 
that the interaction between the SNR G35.6-0.4 and the molecular gas might be responsible for 
the $\gamma-$ray emission. Additionally they argued that the star formation processes taking place in 
the region, could be an alternative or complementary mechanism to explain the very-high energy 
emission.

In this paper, we present new molecular observations of the dense clump coincident with
the HESS J1858+020 center, carried out with the aim of going deeper in the nature identification 
of the very-high energy emission.

\section{Observations}

The molecular observations were performed on July 14 and 15, 2010 with the 10 m Atacama Submillimeter 
Telescope Experiment (ASTE; \citealt{ezawa04}). We used the CATS345 GHz band receiver, which is a two-single 
band SIS receiver remotely tunable
in the LO frequency range of 324-372 GHz. We simultaneously observed \2 J=3--2 at 345.796 GHz and HCO$^{+}$~J=4--3 at
356.734 GHz, mapping a region of 90\s~$\times$ 90\s~centered at $l =$ 35\fdg577, $b = -$ 0\fdg578 
(RA $=$ 18\hh 58\mm 19.5\ss, dec. $=$ $+$02\d 05\m 23.9\s, J2000). Additionally we observed \3 J=3--2 at 330.588 GHz 
and CS J=7--6 at 342.883 GHz towards the same center and mapping a region of 40\s~$\times$ 50\s. The mapping grid 
spacing was 10\s~and the integration time was 60 sec. per pointing in both cases.
All the observations were performed in position 
switching mode. The off position ($l =$ 35\fdg478, $b = -$ 0\fdg540) was checked to be free of emission.

We used the XF digital spectrometer with a bandwidth and spectral resolution set to 128 MHz and 125 kHz, respectively.
The velocity resolution was 0.11 \ks and the half-power beamwidth (HPBW) was 22\s~at 345 GHz. The system temperature
varied from T$_{\rm sys} = 150$ to 200 K. The main beam efficiency was $\eta_{\rm mb} \sim 0.65$.
The spectra were Hanning smoothed to improve the signal-to-noise ratio and only linear or/and some third order
polynomia were used for baseline fitting.
The data were reduced with NEWSTAR and the spectra processed using the XSpec software package.

To complement the new molecular data, we used the mosaiced images from 
GLIMPSE and MIPSGAL surveys from the {\it Spitzer}-IRAC (3.6, 4.5, 5.8 and 8 $\mu$m) and {\it Spitzer}-MIPS 
(24 and 70 $\mu$m), respectively. IRAC has an angular resolution between 1\farcs5 and 1\farcs9 and
MIPS 6\s~at 24 $\mu$m.
Additionally we analyzed the continuum emission at 1.1 mm obtained from the Bolocam Galactic Plane Survey (BGPS) which 
has a FWHM effective resolution of 30\s.

\section{The studied region}
\label{present}

\begin{figure*}[h]
\centering
\includegraphics[width=7.1cm]{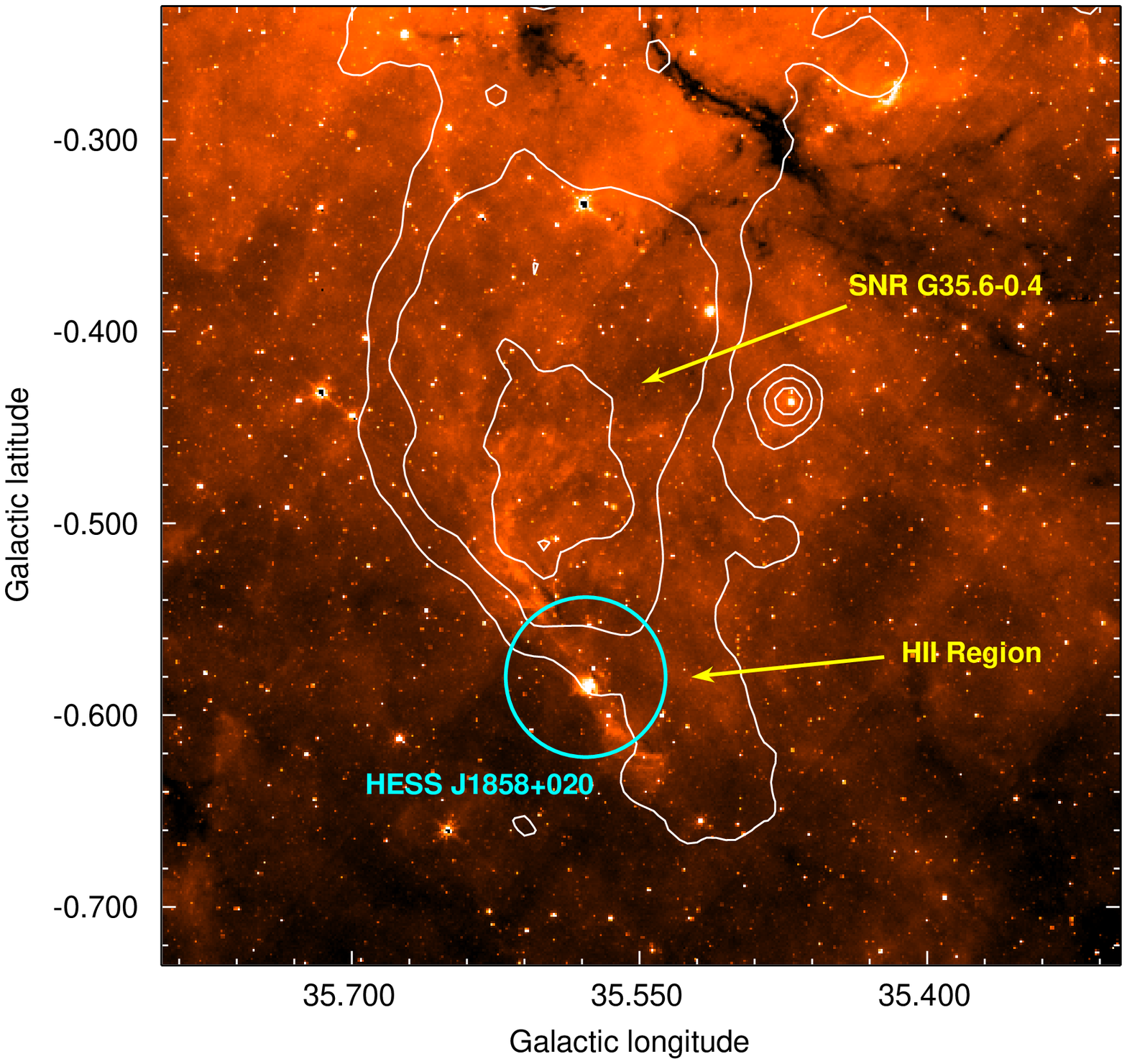}
\includegraphics[width=6.5cm]{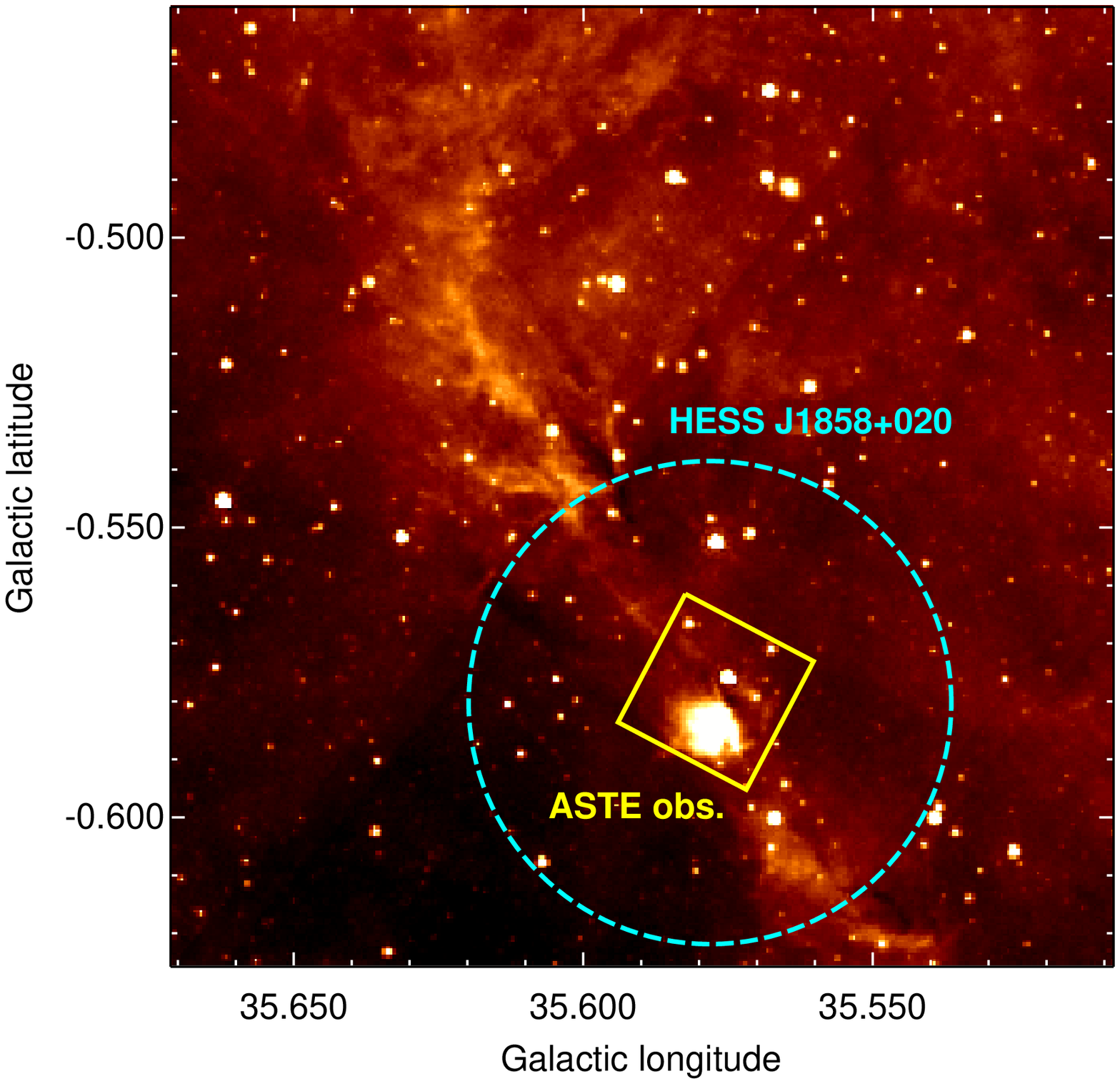}
\caption{Left: region of about 30\m~$\times$ 30\m~towards SNR G35.6-0.4 presenting the emission at 8 $\mu$m with 
contours of the radio continuum emission at 20 cm. The contours levels are 17, 22, and 30 K. The first contour 
is slightly above the data 3$\sigma_{rms}$. The circle shows 
the position and the extension of HESS J1858+020. We remark the possibility that the SNR is partially superimposed 
over an HII region. Right: smaller portion of the region displaying the 8 $\mu$m emission and showing the area mapped 
with the molecular observations (yellow box).}
\label{region}
\end{figure*}

In Figure \ref{region} (left), we present a region of about 30\m~$\times$ 30\m~towards SNR G35.6-0.4. The image 
displays the 8 $\mu$m emission from {\it Spitzer}-IRAC with contours of the radio continuum emission at 20 cm. 
The circle shows the position and the extension of $\sim$5\m~of the source HESS J1858+020 \citep{aha08}. 
Based on the 8 $\mu$m emission, which traces the presence of 
polycyclic aromatic hydrocarbons (PAHs), partially 
bordering the radio continuum emission extending to the south, we suggest that the SNR G35.6-0.4 
partially overlaps an extended HII region, likely to be part of the same complex. This fact probably 
explains the confusion about the nature of G35.6-0.4 in the past years (see \citealt{green09} and 
references therein). Towards the center of HESS J1858+020 there is an emission peak of 8 $\mu$m which, 
as studied by \citet{yo10}, coincides with a molecular clump 
detected in the \3 J=1--0 line. \citet{yo10} have shown evidence 
of star forming activity in coincidence with this clump. This region is catalogued in the IRAS Catalogue 
of Point Sources (Version 2.0; \citealt{iras}) as IRAS 18558+0201. 
Figure \ref{region} (right) shows an enlargement of the area of interest indicating with a yellow box the 
region where the new molecular observations were carried out.

\section{Results and discussion}

\begin{figure}[h]
\centering
\includegraphics[width=9cm]{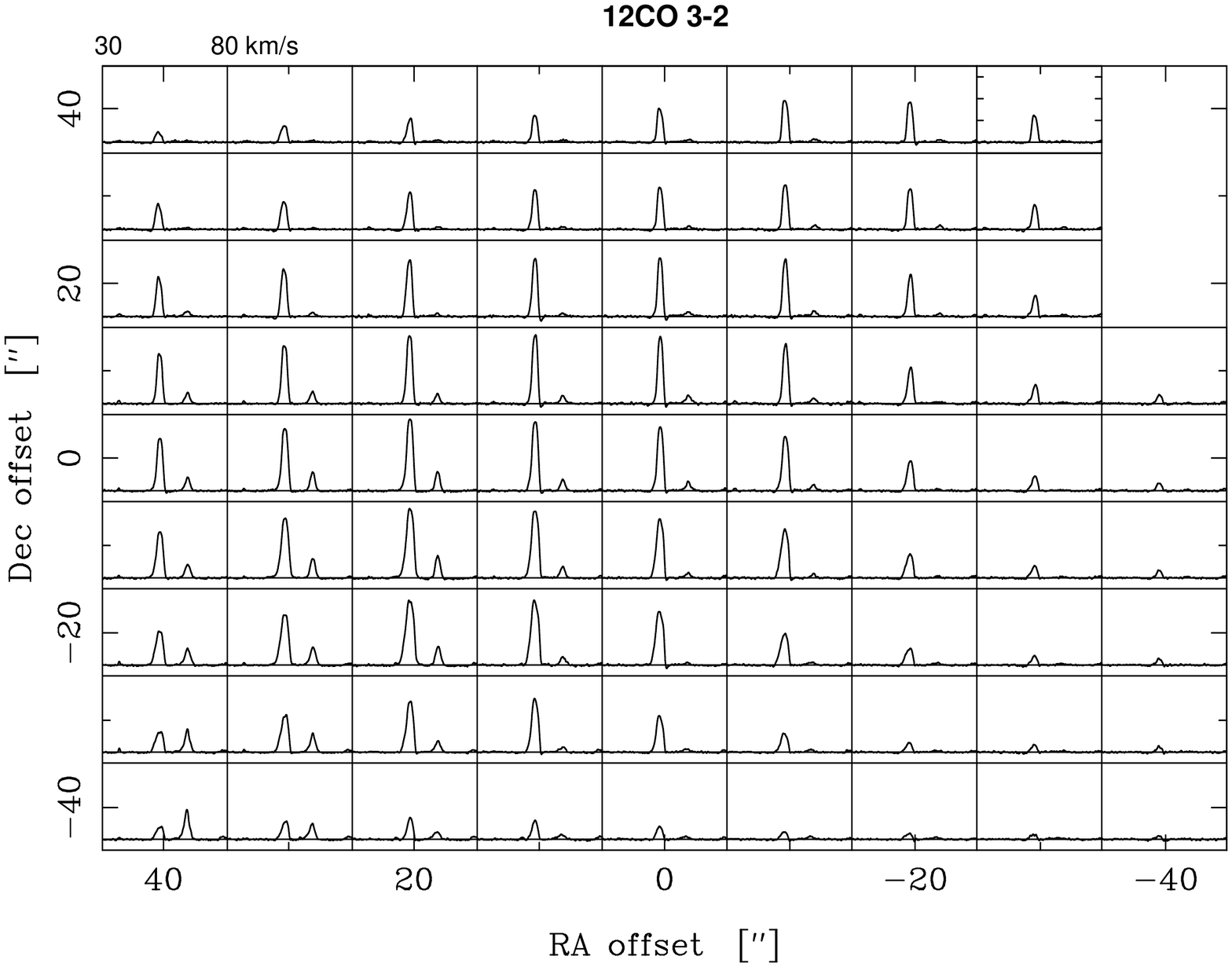}
\includegraphics[width=7.5cm]{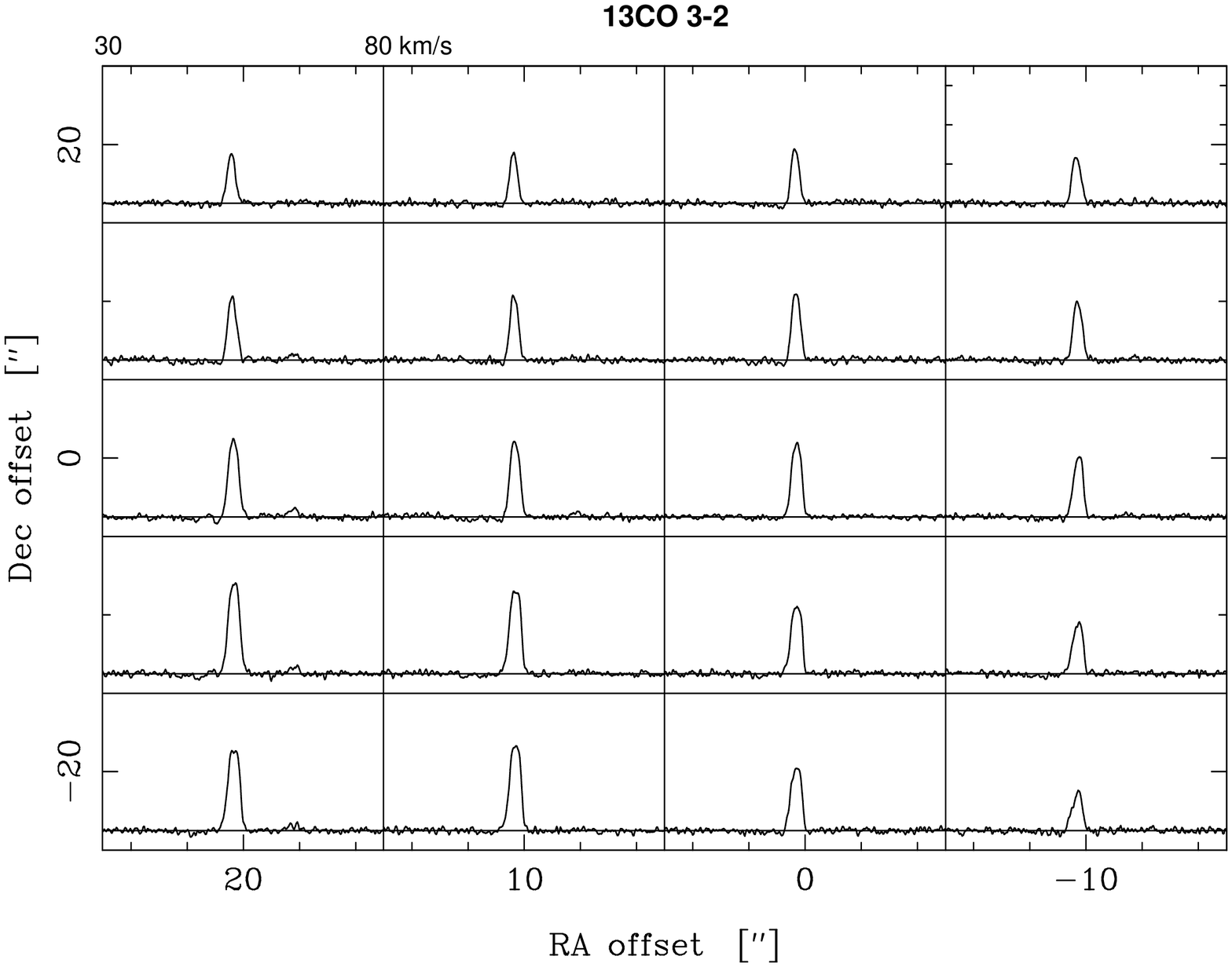}
\caption{Up: \2 J=3--2 spectra. Bottom: \3 J=3--2 spectra. The horizontal axis of each spectra is velocity,
and ranges from 30 to 80 \k, while the vertical axis is brightness temperature and goes from -1 to 7 K. The center,
i.e. the (0,0) offset, in both lines is the same.}
\label{spectr}
\end{figure}

Figure \ref{spectr} (up) shows the \2 J=3--2 spectra obtained towards the observed region. In the whole area 
the main component at $\sim$53 \k, already detected in the \3 J=1--0 clump studied
by \citet{yo10}, is present. A second, less intense, component is observed mainly towards 
positive RA and negative Dec offsets (bottom left in the image) with a velocity of $\sim$64 \k. Note that,
due to lack of observing time, 
three positions were not observed (top right of the image). Fig. \ref{spectr} (bottom) displays the \3 J=3--2 
spectra observed towards the central $\sim$20 square arcseconds. In the observed area, this line has 
a unique component centered at $\sim$53 \k. In both 
cases, the horizontal axis of each spectra is velocity and ranges from 30 to 80 \k, while the vertical
axis is brightness temperature and goes from -1 to 7 K. 
The \2 J=3--2 component at $\sim$64 \ks has no correspondence neither in the \3 J=3--2 emission presented in 
this work, nor in the \3 J=1--0 emission analyzed in \citet{yo10}. We suggest that this velocity component can be 
unrelated molecular gas seen along the line of sight. In what follows, 
we focus our analysis on the $\sim$53 \ks molecular component. 
Table \ref{paramtable} summarizes the derived
parameters of the \2 and \3 J=3--2 lines obtained from a Gaussian fitting. T$_{mb}$ is the main beam peak brightness temperature,
V$_{{\rm LSR}}$ is the central velocity referred to the Local Standard of Rest and $\Delta v$ is the line width (FWHM). 
The Gaussian fitting was performed to
the averaged spectrum of each line, which was obtained from the pointings within the area mapped by the \3 emission, 
at the center of the region. The quoted uncertainties are formal 1$\sigma$ value for the model of the Gaussian shape.

\begin{table}[h]
\caption{Parameters of the \2 and \3 J=3--2 lines from the center of the region.}
\label{paramtable}
\centering
\begin{tabular}{cccc}
\hline\hline
Emission & T$_{mb}$ & V$_{{\rm LSR}}$ & $\Delta v$  \\
         &  (K)     & (\k)      &   (\k)      \\
\hline
\2 J=3--2 & 10.50 $\pm$ 0.40   &  53.15 $\pm$ 0.22  & 2.60 $\pm$ 0.15 \\
\hline
\3 J=3--2 &  5.20 $\pm$ 0.50  &  53.30  $\pm$ 0.15 & 2.00 $\pm$ 0.10  \\
\hline
\end{tabular}
\end{table}

\begin{figure}[h]
\centering
\includegraphics[width=7cm]{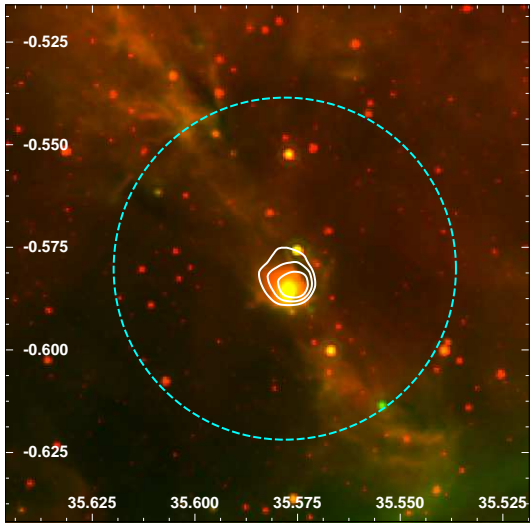}
\caption{Two color image with the 8 $\mu$m and 24 $\mu$m emissions presented in red and green, respectively. The 
contours correspond to the \2 J=3--2 emission integrated between 48 and 57 \k, at the levels of 22, 26, and 30 K \k.
The rms noise is about 4 K \k. The circle represents the extension of the source HESS J1858+020.}
\label{integ}
\end{figure}

An inspection of the \2 J=3--2 spectra indicates that there are not spectral wings nor intensity gradients along symmetric directions 
in the plane of the sky, which allows us to conclude that, at the present data resolution, there is not evidence of outflow activity 
neither in the plane of the sky nor along the line of sight. The detected molecular clump peaks approximately 
at the (10,0) offset (see Fig. \ref{spectr}), corresponding to the sky positon $l =$ 35\fdg57, $b = -$0\fdg58.
Figure \ref{integ} displays a two color image with the 8 $\mu$m and 24 $\mu$m emissions in red and green, respectively, with contours 
of the \2 J=3--2 emission integrated between 48 and 57 \k. The circle represents the source HESS J1858+020. From 
this image it can be appreciated that the molecular clump mapped in \2 J=3--2 coincides with the condensation of PAHs 
seen at 8 $\mu$m. This image reveals that such clump also emits at 24 $\mu$m indicating warm dust. 
The presence of this clump, lying exactly at the geometric center of the HESS source, suggests that its study may 
help to elucidate the nature of the high energy emission.
It is important to note that we did not detect emission of the HCO$^{+}$~J=4--3 and CS J=7--6 lines at a sensitivity levels 
of about 0.13 and 0.2 K, respectively in the direction to this molecular concentration.

To estimate the physical parameters of the molecular clump we assume LTE conditions and a beam filling 
factor of 1, which may not be completely true but allow us to make a first approach to the problem.
From the peak temperature ratio between the CO isotopes ($^{12}$T$_{mb}$/$^{13}$T$_{mb}$), it is possible to estimate 
the optical depths from (e.g. \citealt{curtis10}): 
$$ \frac{^{12}{\rm T}_{mb}}{^{13}{\rm T}_{mb}} = \left(\frac{\nu_{12}}{\nu_{13}}\right)^{2} \frac{[^{12}\rm{CO}]}{[^{13}\rm{CO}]} \frac{1-exp(-\tau_{12})}{\tau_{12}},$$ where $\nu_{12} = 345.796$ GHz and $\nu_{13} = 330.558$ GHz are the 
transition frequencies of \2 and \3 J=3--2 lines, respectively, $\tau_{12}$ is the optical depth of the \2 gas and 
[\2]/[\3] is the isotope abundance ratio. Assuming 8 kpc as the distance to the galactic center and 
using [\2]/[\3] $= (6.21\pm1.00){\rm D_{GC}} + (18.71\pm7.37$)  \citep{milam05} where D$_{\rm GC} = 6.73$ kpc is the distance between 
the source and the galactic center, we obtain [\2]/[\3] $ = 56.7 \pm 13.5$. Thus, the \2 J=3--2 optical depth 
is $\tau_{12} = 32 \pm 11$.
Using the typical LTE equations and taking into account that the \2 J=3--2 line is optically thick as shown above,
from its emission we estimate an excitation temperature of T$_{\rm ex} = 17 \pm 1$ K. 
Using this factor and the \3 J=3--2 emission,
we derive an optical depth for the \3 of $\tau^{13} = 0.70 \pm 0.12 $ and a \3 column density of 
N(\3) $= (8.2 \pm 1.2) \times 10^{15}$ cm$^{-2}$. Adopting the isotope abundance ratio [\2]/[\3] used above and 
the relationship of N(H$_2$)/N(\2) $\sim 10^4$ (see \citealt{black84} and reference therein), we obtain an H$_2$
column density of N(H$_2$) $= (5.0 \pm 1.8) \times 10^{21}$ cm$^{-2}$.
Finally, assuming spherical geometry for the clump, compatible with what is seen in Fig. \ref{integ}, with a radius of 
$\sim$30\s~($\sim$1.5 pc at the distance of 10.5 kpc), we estimate a mass and a volume density of 
$(1.5 \pm 0.5) \times 10^3 \times (\frac{d}{10.5~{\rm kpc}})^2$ \msol, and $(2.1 \pm 0.7) \times 10^{3} \times (\frac{10.5~{\rm kpc}}{d})^3$ cm$^{-3}$, respectively for this structure, where $d$ is the distance. The quoted errors, of the order of 30\%, do not include 
the error in the distance, which is a major unknown and depends on Galaxy models. We thus present the estimated values as a function
of the distance.

On the other hand, using the \3 line width of $\Delta v = 2$ \ks and a radius of $R = 1.5$ pc, we calculate the virial mass
from: M$_{vir} = B \times R \times \Delta v^{2}$, where $B$ is a constant that depends on the density profile. If one assumes 
a uniform density profile, that is $\rho(r) = constant$, $B = 210$, while if a density profile $\rho(r) \propto 1/r$ is assumed, $B = 190$ 
\citep{mac88}. Both cases produce the same virial mass within the errors, M$_{vir} = (1.1 \pm 0.1) \times 10^{3} \times \frac{d}{10.5~{\rm kpc}}$ \msol. 
The ratio between the virial and the LTE mass is M$_{vir}$/M$_{\rm{LTE}} = (0.8 \pm 0.3) \times \frac{10.5~{\rm kpc}}{d}$. 
\citet{kawa98} performed a large scale survey of molecular clouds towards the Gemini and Auriga regions and shows that star-forming \3 clouds have low M$_{vir}$/M$_{\rm{LTE}}$, while all the clouds with high M$_{vir}$/M$_{\rm{LTE}}$ have no sign of star formation. 
On the other hando, several molecular cores studied in the active stellar forming complex in Taurus have on average, M$_{vir}$/M$_{\rm{LTE}} \sim 0.6$ \citep{onishi96}, quite similar to the 0.8 derived here. 

\begin{figure}[h]
\centering
\includegraphics[width=7.3cm]{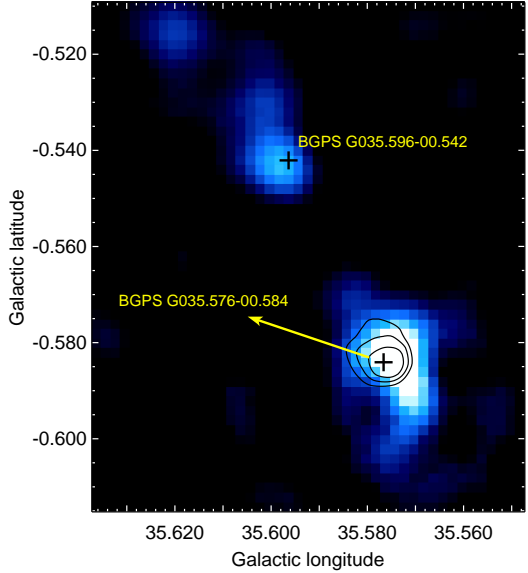}
\caption{BGPS continuum emission at 1.1 mm with the \2 J=3--2 contours presented in Fig. \ref{integ}. 
The crosses indicate the position of two sources catalogued in the BGPS Catalog.}
\label{bolocam}
\end{figure}

Another way to estimate the volume density of the molecular feature is through the investigation 
of the dust content. Figure \ref{bolocam} displays the smoothed 1.1 mm continuum emission obtained from 
the BGPS \citep{aguirre11} with 
the \2 J=3--2 contours presented in Fig. \ref{integ}. The crosses indicate the position of two 
sources from the BGPS catalog \citep{rosolow10}, showing that the analyzed molecular clump coincides with the source
BGPS G035.578-00.584.

According to the BGPS Catalog, the source BGPS G035.578-00.584 has an integrated flux density at 
1.1 mm of $S_{\nu} = 0.37 \pm 0.10$ Jy and an elliptical shape with a major and minor axis of 16\farcs9 and 13\farcs6, 
respectively. The total 
mass of gas and dust in a core is proportional to the total flux density $S_{\nu}$, assuming that the dust 
emission at 1.1 mm is optically thin and 
both the dust temperature and opacity are independent of position within the core \citep{enoch06}. We can then calculate
the core mass from the BGPS G035.578-00.584 flux density using:
$$ M = \frac{d^{2} S_{\nu}}{B_{\nu}(T_{d})\kappa_{\nu}},$$ 
where $\kappa_{1.1{\rm mm}} = 0.0114$ cm$^{2}$ g$^{-1}$ is the dust opacity estimated on the basis of the canonical gas-to-dust
mass ratio of 100 \citep{enoch06}, $d$ the distance, $B_{\nu}$ the Planck
function, and  $T_{d}$ the dust temperature. Although the millimeter emission arises only from the dust, it is 
possible to infer the total mass of gas and dust because, as mentioned above, the dust opacity $\kappa_{1.1{\rm mm}}$ 
contains the gas-to-dust mass ratio (see \citealt{enoch06}). Following \citet{rosolow10}, the above equation can be written as:
$$ M = 13.1 {\rm M_{\odot}} \left(\frac{d}{1 {\rm kpc}}\right)^{2} \left(\frac{S_{\nu}}{1 {\rm Jy}}\right) \left[\frac{exp(13/T_{d}) - 1}{exp(13/20) - 1}\right].$$
Assuming a typical dust temperature of $T_{d} = 20$ K and the distance of 10.5 kpc, we obtain a total mass for the 
core BGPS G035.578-00.584 of $M = (535 \pm 15) \times (\frac{d}{10.5~{\rm kpc}})^2$ \msol.
Finally, using this mass and assuming an angular radius of 14\s~($R \sim 0.7$ pc), from 
$ n_{\rm{H}} = 3M/(4 \pi R^{3} \mu m_{\rm{H}})$, where $ m_{\rm{H}}$ is the hydrogen atom mass 
and $\mu = 2.37$ the mean particle mass, we obtain a particle density of $ n_{\rm{H}} = (6.6 \pm 0.2) \times 10^{3} \times 
(\frac{10.5~{\rm kpc}}{d})^3$ cm$^{-3}$.
The adopted angular radius is based on the angular size of this object as reported in the catalog. As it can be 
appreciated, the minor axis size is smaller than the rms size of the BGPS beam, thus it is not possible to calculate the 
deconvolved angular radius following \citet{rosolow10}.

In summary, from two different methods we obtain a density of a few $10^{3}$ cm$^{-3}$ for the studied clump. Taking
into account the lack of emission of the CS J=7--6 and HCO$^{+}$~J=4--3 lines, tracers of higher densities, 
we conclude that $10^{3} - 10^{4}$ cm$^{-3}$ is a plausible range for the density in the clump.

\subsection{Young stellar objects in the molecular clump}

\citet{yo10} conducted a search for young stellar objects (YSOs) probably embedded in the molecular cloud mapped in 
the \3 J=1--0 line. Using the color criteria of \citet{all04} for GLIMPSE sources, the authors found six 
YSO candidate sources. In this work, we look for YSOs probably embedded in the discovered \2 J=3--2 clump using 
criteria based on the intrinsic reddening of the sources and studying the physical parameters extracted
from the YSOs spectral energy distributions (SEDs).
These criteria are based in the fact that YSOs always show intrinsic infrared excess that cannot be attributed to
scattering and/or absorption of the ISM along the line of sight. We therefore used the GLIMPSE Point-Source Catalog 
to search for this kind of sources within the molecular clump.
\citet{robi08} defined a color criterion to identify intrinsically red sources using data from
the {\it Spitzer}-IRAC bands.
Intrinsically red sources satisfy the condition  $m_{4.5}-m_{8.0}\geq1$, where $m_{4.5}$ and $m_{8.0}$ are
the magnitudes in the 4.5 and 8.0 $\mu$m bands, respectively.
To consider the errors in the magnitudes, we use the following color criterion to select
intrinsically red sources: $m_{4.5}-m_{8.0}+\varepsilon\geq1$, where
$\varepsilon=\sqrt{(\Delta_{4.5})^2+(\Delta_{8.0})^2}$ and
$\Delta_{4.5}$ and $\Delta_{8.0}$ are the errors of the 4.5 and 8.0 $\mu$m bands, respectively.
By inspecting a circular region of about 30\s~in radius centered at $l =$ 35\fdg578, $b = -$0\fdg582, we 
find 27 GLIMPSE sources, and by applying the above mentioned color criterion, we find only two intrinsically red 
sources that appear to be related to the molecular clump, they are: SSTGLMC G035.5768-00.5862 and 
SSTGLMC G035.5765-00.5909, hereafter IRS1 and IRS2, respectively. These sources were classified as class I and II, respectively in
\citet{yo10} following the \citet{all04} classification.
Now, in view of our more complete study, we suggest that IRS1 is very likely embedded 
in the analyzed molecular clump, while for IRS2, lying in the border of the observed region, the connection
with the studied molecular feature is less compelling.

\begin{table}[h]
\caption{Near- and mid-IR fluxes of IRS1 and IRS2.}
\label{irs}
\centering
\begin{tabular}{ccccccccc}
\hline\hline
Source &  $J$ & $H$ & $K_{S}$ &  3.6 $\mu$m & 4.5 $\mu$m  &  5.8 $\mu$m &  8.0 $\mu$m & 24 $\mu$m \\
       & (mag) & (mag) & (mag) & (mag) & (mag) & (mag) & (mag) & (Jy) \\
\hline
IRS1     & -- & -- & 13.462 & 10.253 & 9.225 & 8.301 & 7.533 & 0.032  \\
IRS2     & 15.318 & 13.344 & 12.342 & 10.731 & 10.137 & 9.625 & 9.053 & 0.150  \\
\hline
\end{tabular}
\end{table}

\begin{figure}[h]
\centering
\includegraphics[width=12cm]{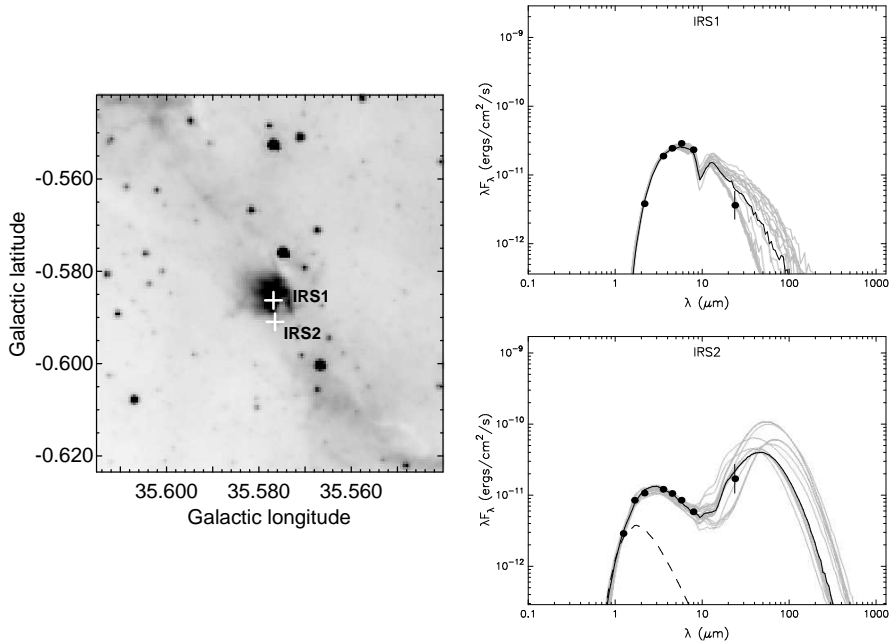}
\caption{Left: Location of sources IRS1 and IRS2. Right: SED of these sources performed
with the tool developed by \citet{robi06,robi07}. Black line shows the best fit, and the
gray lines show subsequent good fits. The dashed line shows the stellar photosphere corresponding to the central
source of the best fit model, as it would look in the absence of circumstellar dust. The points are the
input fluxes.}   
\label{seds}
\end{figure}

In Table \ref{irs} we present the catalogued 
near- and mid-IR fluxes of these sources extracted from the 2MASS and GLIMPSE point source catalogs. 
The fluxes at 24 $\mu$m were obtained from the MIPS image.
These fluxes were used to calculate the SED of IRS1 and IRS2 using the tool developed by \citet{robi06,robi07} and available online\footnote{http://caravan.astro.wisc.edu/protostars/}.
To perform the SED we assume an interstellar absorption between 10 and 30 mag. 
The lower value is compatible with the rough assumption of 1 mag per kpc that is commonly used. 
The upper value is in agreement with the visual absorption obtained from $A_{v}=5\times{10^{-22}}N(H)$ \citep{bohlin78},
where N(H) $=$ N(HI)$+$2N(H$_{2}$) is the line-of-sight hydrogen column density towards this region, which
we estimate in about 5.8 $\times 10^{22}$ cm$^{-2}$. This value was obtained from the HI column density derived from the 
VLA Galactic Plane Survey (VGPS) HI data \citep{stil06} and from the H$_{2}$ column density derived above. 
In Table \ref{tablesed}, we report the main results of the SED fit output for IRS1 and IRS2. In Col. 2 and 3 we report 
the $\chi^{2}$ of the YSO and stellar photosphere best-fit model, respectively.
The remaining columns report the physical parameters of the source for the best-fit model:
central source mass, disk mass, envelope mass, and envelope accretion rate, respectively.
Figure \ref{seds}-right shows the SED of these sources.

\begin{table}[h]
\caption{Parameters derived from the SED fitting of sources IRS1 and IRS2.}
\label{tablesed}
\centering
\tiny
\begin{tabular}{ccccccc}
\hline\hline
Source & $\chi^{2}_{YSO}$ & $\chi^{2}_{\star}$ & M$_{\star}$   & M$_{disk}$    &  M$_{env}$    & $\dot{M}_{env}$  \\
       &                          &                            & (M$_{\odot}$) & (M$_{\odot}$) & (M$_{\odot}$) & (M$_{\odot}/yr$)  \\
\hline
IRS1   & 4.1  & 412 & 9.6  & 3.2 $\times 10^{-2}$ & 3.6 $\times 10^{-6}$ &   0                 \\
IRS2   & 4.6  & 166 & 6.4  & 3.3 $\times 10^{-2}$ & 1.7                  &   5 $\times 10^{-6}$ \\
\hline
\end{tabular}
\end{table}

To relate the SED to the evolutionary stage of the YSO, \citet{robi06} defined three different 
stages based on the values of the central source mass M$_{\star}$, the disk mass M$_{disk}$, 
and the envelope accretion rate $\dot{\rm M}_{env}$ of the YSO. Stage I YSOs
are those that have $\dot{\rm M}_{env}$$/$M$_{\star} > 10^{-6}$ yr$^{-1}$, i.e., protostars with large 
accretion envelopes; stage II are those with M$_{disk}/$M$_{\star} > 10^{-6}$ and $\dot{\rm M}_{env}$$/$M$_{\star} < 
10^{-6}$ yr$^{-1}$, i.e., young
objects with prominent disks; and stage III are those with M$_{disk}/$M$_{\star} < 10^{-6}$ and
$\dot{\rm M}_{env}$$/$M$_{\star} <  10^{-6}$ yr$^{-1}$, i.e. evolved sources where the flux is dominated by 
the central source. According to this classification, we conclude that IRS1 and IRS2 are stage II sources. However, 
IRS2 must be younger than IRS1 because, as can be appreciated in Table \ref{tablesed}, 
it has a massive envelope which must still be accreting mass. The evolutionary stage of IRS1 derived from its SED and the 
lack of evidence of molecular outflows in the clump where it is embedded, suggest that IRS1 is an 
evolved YSO probably in the last stages of formation. Moreover, the presence
of a condensation of PAH around this source suggests that IRS1 could be a high mass protostellar object (HMPO) that 
has not yet reached the ultracompact HII region stage.

\subsection{The scenario}

Based on the results presented above, in this section we discuss about the possible origin of the very-high energy emission. 

As mentioned in Sec. \ref{present} we propose that the SNR G35.6-0.4 is partially overlapping an 
extended HII region, whose eastern border is delineated by PAHs revealed through the 8 $\mu$m emission. A molecular cloud composed
by at least two clumps lies over this border, and one of them is located at the center of HESS J1858+02.
We have shown that there is at least one YSO embedded in such clump (that we called IRS1). Such object can, in principle, 
creates a population of ralativistic particles inside the host molecular cloud via a thermal jet. These particles, in a high density 
ambient matter can originate $\gamma-$ray emission through Inverse Compton and relativistic Bremsstrahlung losses \citep{araudo07}.
However, in the case of IRS1, at the present data resolution, no evidence of molecular outflows has been found neither in the plane
of the sky nor along the line of sight, therefore weakening the probability of a physical link between IRS1 and HESS J1858+020.
Having demonstrated that the presence of a YSO in the molecular clump does not play a decisive role in the production of the 
observed $\gamma-$rays, and in view of the lack of any other candidate in the region at any distance, we are led to the conslusion that the only possible Galactic counterpart for the HESS source is the SNR G35.6-0.4 with its molecular
enviroment. In this case, the supernova shock is a source of accelerated cosmic rays and the dense molecular clump provides the 
nuclei responsible for the production of neutral pions (through inelastic pp collisions) which will decay yielding the observed 
$\gamma-$rays.

\section{Summary }

Using molecular observations obtained with the Atacama Submillimeter Telescope Experiment (ASTE) and IR and submillimeter
continuum archival data, we study a molecular clump associated with the IR source IRAS 18558+0201 that lies at the center of the very-high energy source HESS J1858+020. 
The main results can be summarized as follows:

(a) From the \2 and \3 J=3--2 lines and the 1.1 mm continuum emission we derived for this clump a density between
$10{^3}$ and $10{^4}$ cm$^{-3}$. This clump is part of a larger molecular cloud that is 
being disturbed by the SNR G35.6-0.4 and by a nearby extended HII region.

(b) From the analysis of the mid-IR data and a photometric study we discovered a YSO very likely embedded 
in the mentioned molecular clump. Analyzing its spectral energy distribution we suggest that this source
could be a high mass protostellar object that has not reached the ultracompact HII region stage.

(c) We did not find any evidence of molecular outflows from the discovered YSO that would reveal the presence of 
a thermal jet capable by itself of generating the very-high energy emission.

(d) We conclude that the presence of a clumpy molecular cloud, like the one investigated
in this work, is the most plausible 
explanation for the very-high energy emission. The molecular gas may be acting as a target for the cosmic rays accelerated 
by the shock front of the SNR G35.6-0.4 generating the $\gamma-$ray emission through hadronic processes.

\begin{acknowledgements}

S.P., E.G. and G.D. are members of the {\sl Carrera del
Investigador Cient\'\i fico} of CONICET, Argentina. 
This work was partially supported by Argentina grants awarded by Universidad de Buenos Aires, CONICET and ANPCYT.
M.R wishes to acknowledge support from FONDECYT (CHILE) grant No108033.
She is supported by the Chilean {\sl Center for Astrophysics} FONDAP No.
15010003. S.P. and M.R. are grateful to Dr. Shinya Komugi for the support received during the observations.
We wish to thank the anonymous referee whose comments and suggestions have helped to improve the paper.

\end{acknowledgements}

\bibliographystyle{aa}  
\bibliography{bibHess}
\IfFileExists{\jobname.bbl}{}
{\typeout{}
\typeout{****************************************************}
\typeout{****************************************************}
\typeout{** Please run "bibtex \jobname" to optain}
\typeout{** the bibliography and then re-run LaTeX}
\typeout{** twice to fix the references!}
\typeout{****************************************************}
\typeout{****************************************************}
\typeout{}
}

\end{document}